\documentclass[twocolumn,showpacs,pra,superscriptaddress]{revtex4}
\usepackage{amssymb}
\usepackage{graphicx}
\usepackage{dcolumn}
\usepackage{bm}
\usepackage{amsmath}
\usepackage{epsfig}

\begin{document}

\title{Stern-Gerlach deflection of field-free aligned paramagnetic molecules}

\author{E. Gershnabel}
\affiliation{Department of Chemical Physics, The Weizmann Institute
of Science, Rehovot 76100, ISRAEL}
\author{M. Shapiro}
\affiliation{Department of Chemical Physics, The Weizmann Institute
of Science, Rehovot 76100, ISRAEL}
\affiliation{Department of Chemistry, The University of British Columbia, Vancouver, British Columbia V6T1Z1, Canada }
\author{I.Sh. Averbukh}
\affiliation{Department of Chemical Physics, The Weizmann Institute
of Science, Rehovot 76100, ISRAEL}

\begin{abstract}

The effects of laser-induced prealignment on the deflection of paramagnetic molecules by inhomogeneous static magnetic field are studied. Depending on the relevant Hund's coupling case of the molecule, two different effects were identified: either suppression of the deflection by laser pulses (Hund's coupling case (a) molecules, such as $ClO$), or a dramatic reconstruction of the broad distribution of the scattering angles into several narrow peaks (for Hund's coupling case (b) molecules, such as $O_2$ or $NH$). These findings are important for various applications using molecular guiding, focusing and trapping with the help of magnetic fields.

\end{abstract}

\pacs{37.10.Vz, 33.80.-b, 42.65.Re, 37.20.+j, 33.15.Kr }

\maketitle

\section{Introduction} \label{Introduction_magnetic}
Manipulating the translational motion of atoms and molecules by means of inhomogeneous external fields has been studied intensively for many years. Since the pioneering work of Stern and Gerlach that demonstrated  quantization of atomic trajectories in inhomogeneous magnetic field \cite{SG}, the dynamics of many other systems has been studied both in electric and magnetic fields. An important milestone was, for instance,  separation of molecules in different quantum states in order to create a maser, a molecular amplifier of photons \cite{Townes}. Nowadays, the physics of the deflection of atoms and molecules by inhomogeneous fields is as hot as ever, including studies focused on the motion in the static inhomogeneous electric \cite{McCarthy,Benichou,Loesch,Antoine,reduction} and magnetic \cite{McCarthy,Kuebler,even} fields, and even laser fields \cite{Stapelfeldt,Zhao1,Zhao2,Purcell}. In the case of laser deflection,  some novel applications in molecular optics have recently appeared, such as  molecular lens \cite{Stapelfeldt,Zhao1} and molecular prism \cite{Zhao2,Purcell}.
The interaction between a molecule and an external field depends upon the orientation of the molecule. The field-molecule interactions become intensity-dependent for strong enough fields due to the field-induced modification of the molecular angular motion \cite{Zon,Friedrich}. It was lately shown that the intensity-dependent molecular polarizability-anisotropy provides means for tailoring the dipole force felt by molecules in the laser field \cite{Purcell}.  More recently,  a method for controlling the scattering of molecules in external fields by additional ultrashort laser pulses inducing field-free molecular alignment was suggested \cite{gershnabel1,gershnabel5,gershnabel4}.

In this work we return to the basics, and study the prospects of the ultrafast laser control of molecular deflection in the Stern-Gerlach (SG) arrangement. It was shown in the past that molecular scattering in magnetic fields is affected by rotational alignment caused, for example, by collisions in seeded supersonic beams \cite{Aqullanti}.   Here we  demonstrate that this process can be  efficiently and flexibly controlled by  novel ultafast optical tools allowing for preshaping the molecular angular distribution before the molecules enter the SG apparatus. This can be done with the help of numerous recent techniques for laser molecular alignment, which use single or multiple short laser pulses (transform limited, or shaped) to temporarily align molecular axes along certain directions (for introduction to the rich physics of laser molecular alignment, see, e.g. \cite{Zon,Friedrich,Stapelfeldt2,Kumarappan,Stolow,rich}). Short laser pulses excite rotational wavepackets, which results in a considerable transient molecular alignment after the laser pulse is over, i.e., at field-free conditions.  In the present paper, we will consider only molecules with a permanent magnetic dipole moment, i.e., open shell molecules. The open shell molecules are classified into Hund's coupling cases according to their angular momenta coupling \cite{Townes,Carington}. In the Hund's coupling case (a),  the  angular momentum of electrons and their spin are coupled to the internuclear axis, while in the Hund's coupling case (b), the electronic spin and internuclear axis are not strongly coupled. We will consider magnetic deflection of different paramagnetic molecules subject to a short prealigning laser pulse. In the Hund's coupling case (a), the magnetic moment is coupled to the internuclear axis, and by rotating the molecule, one rotates the magnetic moment as well. This  reduces substantially the Zeeman effect and effectively turns off the interaction between the molecule and magnetic field (like a rotating electric dipole that becomes decoupled from a static  electric field \cite{gershnabel5}). In the Hund's coupling case (b), the magnetic moment is barely coupled to the internuclear axis. However, laser-induced molecular rotation creates an effective magnetic field which adds to the SG field and modifies the deflection dynamics. As a result, as we show below, a broad and sparse distribution of the scattering angles of deflected molecules collapses into several narrow peaks with controllable positions.

The paper is organized as  following. In Sec. \ref{General Theory} we  outline the general theorical framefork: first, we  briefly discuss the SG deflection mechanism (Sec. \ref{Stern-Gerlach deflection}), and provide several needed  facts on the laser-induced field-free alignment in Sec. \ref{Prealignment}. Then, the interaction details for the Hund's coupling case (a) and Hund's coupling case (b) are given in Sec. \ref{Hund case A} and \ref{Hund case B}, respectively. Further discussion of the Hund's coupling case (b) and hyperfine structure  appears in the Appendix in Sec. \ref{NH HFS}. In Sec. \ref{Applications to Molecules} we apply the above theoretical tools to the laser-controlled magnetic scattering of  $ClO$ (Sec. \ref{ClO}), $O_2$ (Sec. \ref{$O_2$}) and $NH$ (Sec. \ref{NH}) molecules. Discussion followed by the conclusions, are presented in Sec. \ref{Discussions} and \ref{Summary}, respectively.

\section{General Theory} \label{General Theory}
\subsection{Stern-Gerlach deflection} \label{Stern-Gerlach deflection}

Once a beam of molecules enters into a SG magnetic field, the initial eigenstates of the system adiabatically become $|\Psi_i(B)\rangle$:

\begin{equation}
|\Psi_i(B)\rangle=\sum_{j}a_{j}(B)|\Psi_j\rangle, \label{Hund case b diagonalized}
\end{equation}
where the coefficients $a_{j}(B)$  depend on the magnetic field B as a parameter, and $|\Psi_j\rangle$ is a basis for the free molecule. In this work we consider the magnetic field to be: $\textbf{B}=B(z)\hat{z}$, i.e., pointing in the $z$ direction, with a practically constant gradient along the $z$ direction in the relevant interaction region.

The force acting on the molecule is given by:
\begin{equation}
Force=-\nabla E=-\frac{\partial E}{ \partial B}\frac{\partial B}{ \partial z},
\label{Force}
\end{equation}
where $E$ is the energy of the molecule. The derivative ${\partial E}/{ \partial B}$  may be obtained by means of the Hellman-Feynman theorem, that is, for a system being in the $i$-th energy eigenstate of the system, the force is proportional to:

\begin{eqnarray}
\frac{\partial E_i}{\partial B}&=&\langle\Psi_i(B)|\frac{\partial H}{\partial B}|\Psi_i(B)\rangle\nonumber\\
&=&\langle\Psi_i(B)|\frac{\partial H_z}{\partial B}|\Psi_i(B)\rangle,
\label{Hellman_Feyman}
\end{eqnarray}
where $H_z$ is the Zeeman term of the Hamiltonian. Since $H_z$ is proportional to $B$, we conclude that a molecule in an energy eigenstate will be deflected by a force that is proportional to:

\begin{eqnarray}
{\cal A}_{i}\equiv\langle\Psi_i(B)|\frac{H_z}{B}|\Psi_i(B)\rangle.\label{A_force}
\end{eqnarray}

Eq. \ref{A_force} will allow us to consider the distribution of forces. In order to take into account the absolute amount of deflection, though, one has to consider the field gradient as well (Eq. \ref{Force}). For more details, see, e.g. \cite{McCarthy}.

\subsection{Laser-induced field-free alignment} \label{Prealignment}

If the molecules are subject to a strong linearly polarized femtosecond laser pulse, the corresponding molecule-laser interaction potential is given by:
\begin{equation}
H_{ML}=-\frac{1}{4}\epsilon^2\left [ (\alpha_{\parallel}-\alpha_{\perp}) \cos^2\theta+\alpha_{\perp}\right ],\label{pre alignment interaction}
\end{equation}
where $\theta$ is the angle between the molecular axis and the polarization direction of the pulse, $\alpha_{\parallel},\alpha_{\perp}$ are the parallel and perpendicular polarizability components, and $\epsilon$ is the femtosecond pulse envelope. Since the aligning pulse is short compared to the typical periods of molecular rotation, it may be considered as a delta-pulse. In the impulsive approximation, one obtains the following relationship between the wavefunction before and after the pulse applied at $t=0$ (see e.g. \cite{Gershnabel3}, and references therein):
\begin{equation}
\Psi(t=0^+)=\exp{[iP\cos^2\theta]}\Psi(t=0^-),\label{prealignment operator}
\end{equation}
where the kick strength, $P$ is given by:

\begin{equation}
P=(1/4\hbar)\cdot (\alpha_{\parallel}-\alpha_{\perp})\int_{-\infty}^{\infty}\epsilon^2(t)dt.\label{kick strength}
\end{equation}

We assume the vertical polarization  of the pulse (along the $z$-axis, and parallel to the SG magnetic field). Physically, the dimensionless kick strength $P$, is equal to the typical amount of angular momentum (in the units of $\hbar$) supplied by the pulse to the molecule. In order to find $\Psi(t=0^+)$ for any initial state, we introduce an artificial parameter $\xi$ that will be assigned the value $\xi=1$ at the end of the calculations, and define:

\begin{eqnarray}
\Psi_{\xi}=\exp{\left [ (iP\cos^2\theta)\xi \right ]}\Psi(t=0^-)
=\sum_{i} c_{i}(\xi)|\Psi_i\rangle.\label{artificial}
\end{eqnarray}

By differentiating both sides of Eq. \ref{artificial} with respect to $\xi$, we obtain the following set of differential equations for the coefficients $c_{i}$:

\begin{equation}
\dot{c}_{i'}=iP\sum_{i}c_{i}\langle  \Psi_{i'}|\cos^2\theta|\Psi_i\rangle,
\label{Pre-Alignment Coefficients}
\end{equation}
where $\dot{c}=dc/d\xi$. Evaluation of the matrix elements in Eq. \ref{Pre-Alignment Coefficients} is easily obtained by means of the relationship: $\cos^2\theta=(2D^2_{00}+1)/3$,  where $D^k_{pq}$ is the rotational matrix.

Since $\Psi_{\xi=0}=\xi(t=0^-)$ and $\Psi_{\xi=1}=\Psi(t=0^+)$ (see Eq. \ref{artificial}), we solve numerically this set of equations from $\xi=0$ to $\xi=1$, and find $\Psi(t=0^+)$. It turns out that the population of rotational levels of the kicked molecules has a maximum at around $\hbar P$.

Finally, we derive the distribution of forces acting on a thermal ensemble of molecules pre-aligned by a laser pulse. For this, we start from a single eigenstate of a free system, apply an alignment pulse in the $z$ direction, and then adiabatically increase the magnetic field (in order to imitate a smooth process of  the molecular beam injection  into the SG deflector). The distribution will be proportional to:

\begin{eqnarray}
f({\cal A})&=&\sum_{i,j}\frac{\exp\left(-\frac{E_{i}}{k_B T}\right)}{Q_{rot}}\nonumber\\
&\times&|c_{j}|^2\delta_{{\cal A},{\cal A}_{j}},
\label{distribution of forces}
\end{eqnarray}
where $k_B$ is the Boltzmann's constant, $Q_{rot}$ is the partition function, $i$ denotes the quantum numbers associated with the initial eigenstates of free molecules, $c_{j}$ denotes the coefficients of the free eigenstates that were excited by the laser pulse applied to the initial eigenstate $i$, and ${\cal A}_{j}$ are the associated matrix elements given in Eq. \ref{A_force} (proportional to the force), between the states adiabatically correlated with the free states $j$.

\subsection{Hund's coupling case (a)} \label{Hund case A}

In this subsection we concentrate on the $^{35}ClO$ molecule, which presents a good example for the Hund's coupling case (a). Denoted as $^2\Pi$ in its electronic ground state, it has a nuclear spin $I=3/2$, and it was studied well in the past \cite{Carrington1,Kakar,Brian}.

In the Hund's coupling case (a), the electronic angular momentum and spin are strongly coupled to the internuclear axis, and in the case of $ClO$, its effective Hamiltonian is given by \cite{Carington}:

\begin{equation}
H_{eff}=H_{rso}+H_{hf}+H_Q,\label{ClO molecule Hamiltonian}
\end{equation}
where $H_{rso}$ is the rotation and spin-orbit coupling, $H_{hf}$ is the magnetic hyperfine interaction, and $H_Q$ is the electric quadrupole interaction. Here

\begin{equation}
H_{rso}=B_r\left\{T^1(\textbf{J})-T^1(\textbf{L})-T^1(\textbf{S})\right\}^2+AT^1(\textbf{L})\cdot T^1(\textbf{S}),\label{rso Hamiltonian}
\end{equation}
where $T^1()$ is a spherical tensor of rank $1$, $B_r$ is the rotational constant in the lowest vibrational level, and $A$ is the spin-orbit coupling constant. $\textbf{L}$ and $\textbf{S}$ are the electronic angular momentum and spin operators, respectively. The total angular momentum is $\textbf{J}=\textbf{N}+\textbf{L}+\textbf{S}$, where $\textbf{N}$ is the nuclei angular momentum operator. The Hund's coupling case (a) basis looks like this:
\begin{equation}
|\eta,\Lambda;S,\Sigma;J,\Omega,I,F,M_F\rangle,
\label{Hund case A basis}
\end{equation}
where $\eta$ represents some additional electronic and vibrational quantum numbers, $\Sigma$ and $\Lambda$ are the projections of the electronic spin and angular momentum on the internuclear axis, respectively. For $ClO$ molecule, $S=1/2$, so that $\Sigma=\pm 1/2$ and $\Lambda=1$. The quantity $\Omega$ is $\Omega\equiv\Sigma+\Lambda$ ($\Omega=3/2,1/2$,  the $3/2$-state has a lower energy), and $\textbf{F}=\textbf{J}+\textbf{I}$.

The $H_{hf}$ Hamiltonian is given by:
\begin{eqnarray}
&&H_{hf}=H_{IL}+H_F+H_{dip}\nonumber\\
&=&aT^1(\textbf{I})\cdot T^1(\textbf{L})+b_FT^1(\textbf{I})\cdot T^1(\textbf{S})\nonumber\\
&-&\sqrt{10}g_S\mu_B g_N \mu_N (\mu_0 /4\pi)T^1(\textbf{I})\cdot T^1(\textbf{S},\textbf{C}^2).\label{H_hf_ClO}
\end{eqnarray}

The first term represents the orbital interaction, the second one accounts for the Fermi contact interaction, and the third term describes the dipolar hyperfine interaction. Here $a$ and $b_F$ are constants, $g_N$ and $g_S$ are the nuclear and electron $g$ factors, respectively, $\mu_N$ and $\mu_B$ are the nuclear and electron Bohr magnetons, respectively, and $\mu_0$ is the vacuum permeability.

All the matrix elements for the Hund's coupling case (a), including those for  the quadrupole interaction, are given in \cite{Carington}. The $ClO$ constants were taken from \cite{Carington,Kakar,Brian}.

When considering the Zeeman Hamiltonian, we will concentrate only on the two major terms related with electronic angular momentum and spin:
\begin{equation}
H_Z=\mu_B T^1(\textbf{B})\cdot T^1(\textbf{L})+g_S\mu_B T^1(\textbf{B})\cdot T^1(\textbf{S}).\label{ClO Zeeman}
\end{equation}
The corresponding matrix elements (see Eq. \ref{A_force}) are given in \cite{Carington}.

In order to consider the effect of laser-induced alignment (see Sec. \ref{Prealignment}, Eq. \ref{Pre-Alignment Coefficients}), we have derived the following matrix elements:

\begin{eqnarray}
&\langle& \eta,\Lambda; S,\Sigma;J,\Omega,I,F,M_F|D^{2*}_{00}|\eta,\Lambda;S,\Sigma;J',\Omega,I,F',M_F\rangle\nonumber\\
&=&(-1)^{F-M_F}\left( \begin{array}{ccc} F & 2 & F' \\ -M_F & 0 & M_F \end{array} \right)(-1)^{F'+J+I+2}\nonumber\\
&\times& \sqrt{(2F'+1)(2F+1)}\left\{ \begin{array}{ccc} J' & F' & I \\ F & J & 2 \end{array} \right\}(-1)^{J-\Omega}\nonumber\\
&\times&\left( \begin{array}{ccc} J & 2 & J' \\ -\Omega & 0 & \Omega \end{array} \right)\sqrt{(2J+1)(2J'+1)}.\nonumber\\\label{ClOPulse}
\end{eqnarray}

\subsection{Hund's coupling case (b)} \label{Hund case B}

We will continue by discussing the Hund's coupling case (b), and consider the Oxygen molecule, in its predominant isotopomer $^{16}O^{16}O$. This molecule is probably the most important species among $^3\Sigma$ ground state molecules, and it was one of the first molecules studied in detail \cite{Kuebler,Tinkham}. It is a homonuclear diatomic molecule, where only odd N's appear because of the Pauli's principle and symmetry. This molecule is described well by the Hund's coupling case (b), with the effective Hamiltonian \cite{Carington}:

\begin{equation}
H_{eff}=H_{rot}+H_{ss}+H_{sr}.\label{Effective_Hamiltonian}
\end{equation}

Let us describe separately each term in Eq. \ref{Effective_Hamiltonian}. Here

\begin{equation}
H_{rot}=B_r\textbf{N}^2-D\textbf{N}^4,\label{Rotational}
\end{equation}
is the energy of the nuclei rotation, where $D$ is the centrifugal distortion coefficient. In addition,

\begin{equation}
H_{ss}=-g_s^2\mu_B^2(\mu_0/4\pi)\sqrt{6}T^2(\textbf{C})\cdot T^2(\textbf{S}_1,\textbf{S}_2),\label{SpinSpin}
\end{equation}
is the electornic spin-spin dipolar interaction. $T^2()$ is a spherical tensor of rank $2$. $\textbf{S}_1,\textbf{S}_2$ are electronic spin operators. $T^2_q(\textbf{C})=\langle C^2_q(\theta,\phi)R^{-3}\rangle$, where $C^2_q$ is the spherical harmonics, and

\begin{equation}
H_{sr}=\gamma T^1(\textbf{N})\cdot T^1(\textbf{S}),\label{SpinRotation}
\end{equation}
is the electronic-spin rotation interaction. The Hund's coupling case (b) basis looks like this:
\begin{equation}
|\eta,\Lambda;N,\Lambda;N,S,J,M_J\rangle. \label{Hund case b}
\end{equation}

Here $N$ is the nuclei rotational quantum number, $\Lambda=0$ in our case, $\textbf{S}$ is the electronic spin, which is $1$ in our case, $\textbf{J}=\textbf{N}+\textbf{S}$, and $M_J$ is the projection of $\textbf{J}$ onto a fixed $z$-direction in space. All the needed matrix elements and constants are given in \cite{Carington}. The Zeeman Hamiltonian is given by:

\begin{equation}
H_Z=g_S \mu_B T^1(\textbf{B})\cdot T^1(\textbf{S}).\label{Zeeman Oxygen}
\end{equation}
Its matrix elements (Eq. \ref{A_force}) are given by:

\begin{eqnarray}
&d \langle&  \eta ,\Lambda;N,\Lambda;N,S,J,M_J|T^1_0(\textbf{S})|\eta,\Lambda;N',\Lambda;N',S,J',M_J\rangle\nonumber\\
&=& d (-1)^{J-M_J}\left( \begin{array}{ccc} J & 1 & J' \\ -M_J & 0 & M_J \end{array} \right) \delta_{N,N'}(-1)^{J'+S+1+N}\nonumber\\
&\times&\sqrt{(2J'+1)(2J+1)}\left\{ \begin{array}{ccc} S & J' & N \\ J & S & 1 \end{array} \right\}\nonumber\\
&\times& \sqrt{S(S+1)(2S+1)},
\label{Zeeman Deivation}
\end{eqnarray}
where $d\equiv  g_S \mu_B$.
Finally, in order to account for the laser-induced prealignment, we derived the following relation (to be used in Eq. \ref{Pre-Alignment Coefficients}):

\begin{eqnarray}
&&\langle  \eta ,\Lambda;N,\Lambda;N,S,J,M_J|D^{2*}_{00}|\eta,\Lambda;N',\Lambda;N',S',J',M_J\rangle\nonumber\\
&=& (-1)^{J-M_J}\left( \begin{array}{ccc} J & 2 & J' \\ -M_J & 0 & M_J \end{array} \right) \delta_{S,S'}(-1)^{J'+S+2+N}\nonumber\\
&\times&\sqrt{(2J'+1)(2J+1)}\left\{ \begin{array}{ccc} N' & J' & S \\ J & N & 2 \end{array} \right\}(-1)^{N-\Lambda}\nonumber\\
&\times&\left( \begin{array}{ccc} N & 2 & N' \\ -\Lambda & 0 & \Lambda \end{array} \right)\sqrt{(2N+1)(2N'+1)}.
\label{COS2Derivation}
\end{eqnarray}

In the case of the oxygen molecule, there is a relatively strong effect of the spin-spin interaction, which complicates our analysis. Therefore, we have also chosen an additional $^3\Sigma$ molecule, $^{14}NH$ for our study. For this molecule the ratio between the spin-spin and spin-rotation interactions is reduced (compared to the $O_2$ case). This makes $NH$ a simpler candidate to test our rotational effects. The $NH$ molecule was thoroughly studied in the past \cite{Wayne,Klaus,Jesus,Lewen}, and its effective Hamiltonian is:
\begin{equation}
H_{eff}=H_{rot}+H_{ss}+H_{sr}+H_{HFS}, \label{Hamiltonian_NH}
\end{equation}
where $H_{rot}$, $H_{ss}$ and $H_{sr}$ were defined in Eq. \ref{Rotational}, \ref{SpinSpin} and \ref{SpinRotation}. Since $NH$ has non-zero nuclei spin ($N$ has nuclear spin $I=1$, $H$ has $I=1/2$), then it has a hyperfine structure described by the Hamiltonian $H_{HFS}$. Further elaboration on the hyperfine structure of NH (including details on the Zeeman term, and the matrix elements related to Eq. \ref{Pre-Alignment Coefficients}), is given in the Appendix in Sec. \ref{NH HFS}.

\section{Laser control of the Stern-Gerlach scattering} \label{Applications to Molecules}
\subsection{$ClO$} \label{ClO}

\begin{figure}[htb]
\begin{center}
\includegraphics[width=70mm]{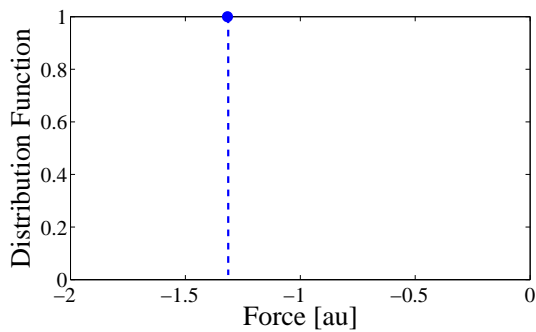}
\end{center}
\caption{The force distribution for a beam of $ClO$ molecules that are deflected by a magnetic field of $1T$. The temperature is $0K$, therefore only the ground state is considered, and the distribution reduces to a single-value peak.}
\label{ClO Distribution no kick}
\end{figure}

In this part of the work, we  apply the theoretical tools that were presented in the previous sections to the SG scattering of the $ClO$ molecule. This molecule exhibits a good Hund's coupling case (a), and details about it were already given in Sec. \ref{Hund case A}. We will consider here its ground state ($T=0K$), for which $\Lambda=1,\Sigma=1/2,\Omega=3/2,J=3/2,F=0,M_F=0$. In Fig. \ref{ClO Distribution no kick} we present the force distribution (Eq. \ref{distribution of forces}) for a $ClO$ molecule in the ground state that is deflected by a SG magnetic field. As only single molecular state is occupied, the force has a well defined single value.

\begin{figure}[htb]
\begin{center}
\includegraphics[width=70mm]{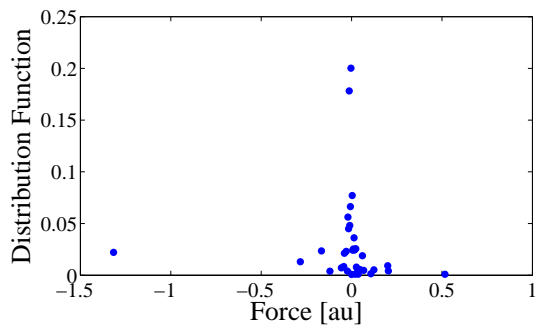}
\end{center}
\caption{The force distribution for a beam of prealigned $ClO$ molecules. The temperature is $0K$, and the kick strength of the laser is $P=30$. The prealigned molecule is deflected by a magnetic field of $1T$. This distribution should be compared to the one from Fig. \ref{ClO Distribution no kick}.}
\label{ClO Distribution P30}
\end{figure}

As the next step, we assume that the molecules are subject to a short laser pulse with a kick strength of $P=30$ (Eq. \ref{kick strength}) before they enter the SG magnetic field. The new force distribution is given in Fig. \ref{ClO Distribution P30}.

By comparing Fig. \ref{ClO Distribution no kick} to Fig. \ref{ClO Distribution P30}, it can be observed that the effect of the laser-induced field-free alignment is to effectively turn-off the interaction between the molecule and the magnetic field. This effect is similar to the one discussed by us recently in connection with the scattering of polar molecules by inhomogeneous static electric fields \cite{gershnabel5}. Moreover,  rotation-induced dispersion in molecular scattering by static electric fields was used as a selection tool in recent experiments on laser molecular alignment \cite{reduction}. A related phenomenon of the reduction of the electric dipole interaction in highly excited stationary molecular rotational states was observed there.  Further details and discussion about the $ClO$ magnetic deflection is provided in Sec. \ref{Discussions}.

\subsection{$O_2$} \label{$O_2$}

\begin{figure}[htb]
\begin{center}
\includegraphics[width=70mm]{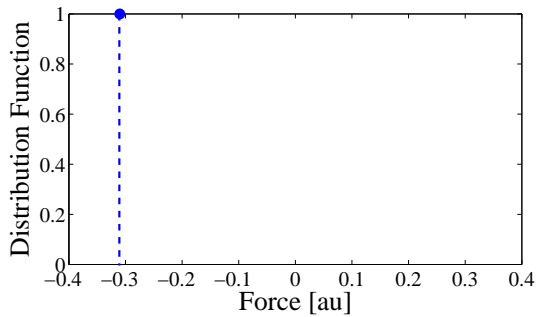}
\end{center}
\caption{The force distribution for a beam of $O_2$ molecules that are deflected by a $1T$ magnetic field. The temperature is $0K$, i.e., only the ground state is populated and therefore the distribution reduces to a single-value peak.}
\label{Results1}
\end{figure}

In this sub-section we consider the $O_2$ molecule. The $O_2$ molecule is described well by Hund's coupling case (b) scheme, and the details about it were given in Sec. \ref{Hund case B}. First, we consider  a beam of $O_2$ molecules at $0K$, i.e., in the ground state ($N=1$, $J=0$, and $M_J=0$). These molecules enter a magnetic field of $1T$, and are deflected by this field. The force distribution for these molecules is given in Fig. \ref{Results1}.

\begin{figure}[htb]
\begin{center}
\includegraphics[width=80mm]{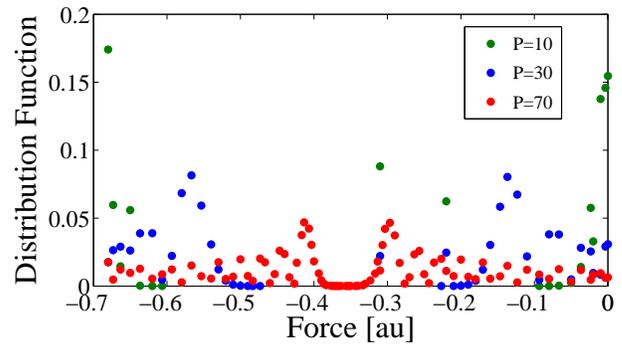}
\end{center}
\caption{The force distribution for a beam of prealigned $O_2$ molecules. Different kick strengths ($P=10,30,70$) are considered and the magnetic field is $1T$ (temperature is $0K$). As the excitation is increased, the two major peaks become closer to each other. This distribution should be compared to the one from Fig. \ref{Results1}.}
\label{FinalPlot1}
\end{figure}

Second, we consider the action of the prealignment pulses of different kick strengths ($P=10, 30, 70$) before the molecules enter the deflecting field. The distribution of forces at $0K$ is given in Fig. \ref{FinalPlot1}, where two major peaks are observed. As the strength of the pulses is increased, higher rotational states are excited, and the peaks become closer to each other.

\begin{figure}[htb]
\begin{center}
\includegraphics[width=80mm]{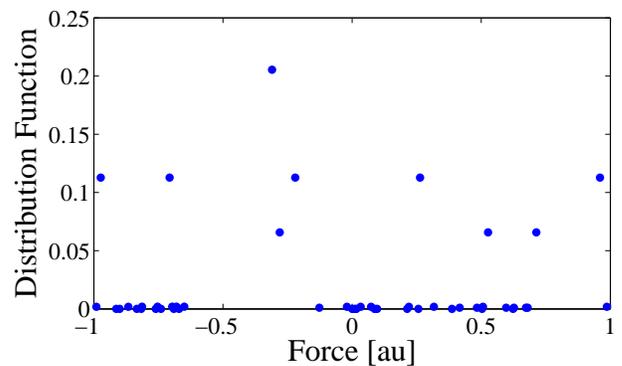}
\end{center}
\caption{The force distribution for $O_2$ molecules. The temperature is $5K$ and the magnetic field is $1T$.}
\label{Results3}
\end{figure}

\begin{figure}[htb]
\begin{center}
\includegraphics[width=60mm]{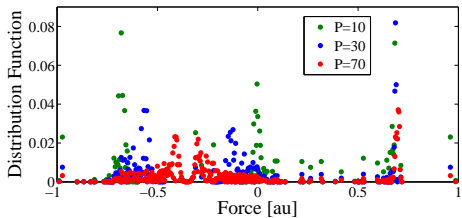}
\end{center}
\caption{The force distribution for a beam of $O_2$ molecules, prealigned by a laser field ($P=10,30,70$). The temperature is $5K$, and the magnetic field is $1T$. Here we observe three major peaks. As the laser excitation strength is increased, the peaks become narrower, and the two left peaks become closer to each other.}
\label{Results4}
\end{figure}

Third, we consider deflection of thermal molecules without and with prealignment, in Figs \ref{Results3} and \ref{Results4}, respectively. By comparing Fig. \ref{FinalPlot1} to Fig. \ref{Results4}, we observe an additional peak in Fig. \ref{Results4}. As the strength of the prealignment pulses is increased, the peaks in Fig. \ref{Results4} are changed: they become narrower and the two left peaks become closer to each other. Further discussion on $O_2$ will be provided in Sec. \ref{Discussions}.

\subsection{$NH$} \label{NH}

\begin{figure}[htb]
\begin{center}
\includegraphics[width=80mm]{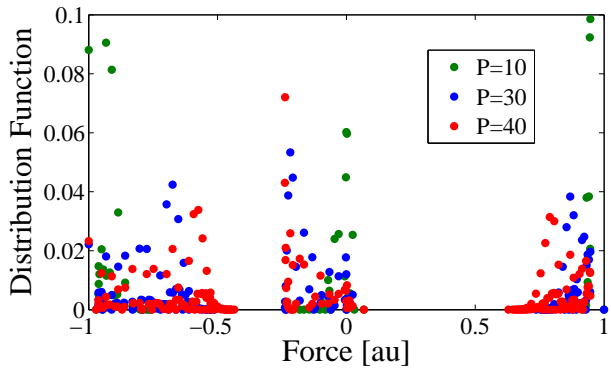}
\end{center}
\caption{The distribution of forces for a beam of $NH$ molecules, that were prealigned (starting from the lowest state $N=0,J=1,F_1=3/2,F=1/2$) by means of laser pulses of different strengths: $P=10$ (green), $P=30$ (blue) and $P=40$ (red). Only $M_F=1/2$ is considered, at a $2T$ magnetic field.}
\label{NH_distribution}
\end{figure}

Finally, we consider the $NH$ molecule. This molecule is described well by a Hund's coupling case (b) scheme, similar to the $O_2$ molecule, however it has a reduced spin-spin to spin-rotation interaction ratio. This makes the $NH$ molecule a simpler candidate for the theoretical analysis.
In Fig. \ref{NH_distribution} we plot the force distribution for the ground state $N=0,J=1,F_1=3/2,F=1/2$ molecules that were prealigned by laser pulses of different intensity. $M_F$ was taken to be $1/2$ for certainty, and the case of $M_F=-1/2$ may be considered similarly (with similar consequences, as will be described in Sec. \ref{Discussions}).

One may observe the presence of three major peaks now (for the $O_2$ molecules in the ground state there were only two peaks). As the strength of the prealignment pulse is increased, the major peaks are shifted in position. An additional difference between the $NH$ and the $O_2$ molecules is that now the peak to the right is also shifted with increasing the strength of the prealignment laser pulse. Further elaboration about this molecule, and the difference between it and $O_2$, is given in Sec. \ref{Discussions}.

\section{Discussion} \label{Discussions}

\subsection{$ClO$} \label{Discussion, ClO}

First we will discuss the $ClO$ molecule, which exhibits a good Hund's coupling case (a). Having both electronic angular momentum and spin  coupled to the internuclear axis, rotation of the molecule by means of short laser pulses leads to the rotation of the molecular magnetic moment  as well. The interaction between the SG magnetic field and the rapidly rotating magnetic moment of the molecule will be thus averaged to zero, leading to the negligible magnetic forces.

\subsection{$O_2$} \label{Discussion, O2}
\begin{figure}[htb]
\begin{center}
\includegraphics[width=60mm]{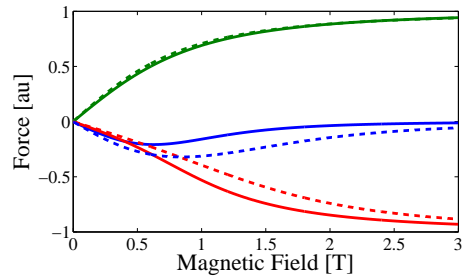}
\end{center}
\caption{Forces vs. magnetic field for the $O_2$ molecule. The $y$ axis is given in arbitrary units, the $x$ axis is given in the units of Tesla. Blue, green, and red (solid lines) correspond to $N=31$, $J=30,31,32$ ($M_J=0$), respectively. Blue, green, and red (dashed lines) correspond to $N=71$ and $J=70,71,72$ ($M_J=0$), respectively. The effects of the spin-spin interaction reveal themselves in the fact that the upper level $J=N$ is well separated from two almost degenerate levels with $J=N+1,N-1$. Magnetic field near $1T$ is  optimal for observing the sensitivity of the deflecting force to the $N$ variation.}
\label{RegularLamda1}
\end{figure}

In Fig. \ref{RegularLamda1} we plot the forces vs. magnetic field, for several values of $J$. First, we observe that for a high magnetic field all the curves are separated to form a three SG splitting pattern \cite{Kuebler}. In the limit of the low magnetic field (and slow rotations), the energy spectrum of the molecule is rather complex due to the spin-spin interaction \cite{Tinkham}. At around $1T$, though, we are in the regime where the spin-rotation ($H_{sr}$) interaction has a rather strong dynamic control: as $N$ is increased (by the means of prealignment, for instance) then a sizable shift of the force magnitude is observed. This is the origin of the behavior of the distributions of Fig. \ref{FinalPlot1} and Fig. \ref{Results4}. It can also be observed in Fig. \ref{RegularLamda1} that the spin-spin term is larger than the spin-rotation one, and it shifts the curve for $J=N$ from the two other curves. We find also that in the case of $J=N+1,N-1$, the forces are more susceptible to different $N$s, which is observed in the distribution of forces in Fig. \ref{FinalPlot1} and Fig. \ref{Results4}.

Fig. \ref{RegularLamda1} also allows us to understand the position of the peaks in Fig. \ref{FinalPlot1} and Fig. \ref{Results4}. The right peak that appears in Fig. \ref{Results4} and does not appear in Fig. \ref{FinalPlot1} corresponds to the $J=N$ states. Due to selection rules (Eq. \ref{COS2Derivation}), the odd $J$s, i.e., the $J=N$ states, are never excited (if we start from $J=0$, and $M_J=0$ at $0K$). This is why we observe only two peaks, i.e., the $J=N\pm1$ peaks, in Fig. \ref{FinalPlot1}. Considering a deflection of the molecules in the ground state alone is important experimentally. Even if one considers an experiment at $T=1K$ ($k_BT=20837MHz$), then the difference between ($N=1, J=0$) and ($N=1, J=2$) (the next energy) is $62486MHz$, which is large enough. Though, for $1K$ we should expect a small peak in the distribution of forces for $J=N$ states. In the case of larger temperature (Fig. \ref{Results4}), we start from different $M$s, and also the odd $J$s are present, therefore, we observe the right peak at Fig \ref{Results4}. As the prealignment becomes stronger, the distribution transforms into three peaks, each correspond to either $J=N$, $J=N-1$ or $J=N+1$ states.

\subsection{$NH$ (and the imaginary $\widetilde{O_2}$ molecule!)} \label{Discussion, NH}

\begin{figure}[htb]
\begin{center}
\includegraphics[width=70mm]{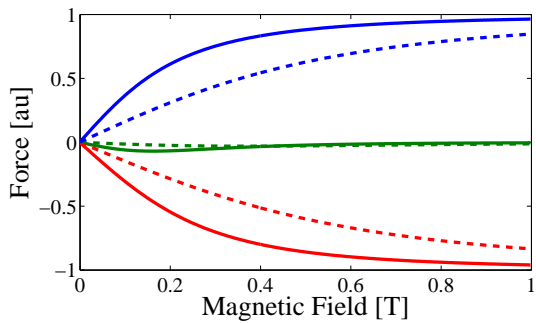}
\end{center}
\caption{Forces vs. magnetic field for the imaginary $\widetilde{O_2}$ molecule (details in the text). The $y$ axis is given in arbitrary units, the $x$ axis is given in the units of Tesla. Blue, green, and red (solid lines) correspond to $N=31$, $J=30,31,32$ ($M_J=0$), respectively. Blue, green, and red (dashed lines) correspond to $N=71$ and $J=70,71,72$ ($M_J=0$), respectively. At about $1T$ for this molecule, we observe approximately a symmetric splitting into three graphs for $J=N$ and $N\pm1$, where $J=N$ is in the middle (unlike in the $O_2$ case). }
\label{LamdaSmall}
\end{figure}

Before we start with the $NH$ molecule, we consider an imaginary $\widetilde{O_2}$ molecule (!). This molecule is similar to the $O_2$ molecule, only with a spin-spin interaction that is reduced by a factor of $100$. The forces vs. magnetic field for the imaginary $\widetilde{O_2}$ molecule are plotted in Fig. \ref{LamdaSmall}, where we get a symmetric splitting of the $N$ level into $J=N,N\pm1$ ($J=N$ is in the middle, as is intuitively expected for SG splittings, unlike in the $O_2$ case). Such behavior corresponds to a spin-spin interaction that is negligible compared with the spin-rotation.

As we intuitively suggested in Sec. \ref{Introduction_magnetic}, when one applies prealignment to molecules belonging to the Hund's coupling case (b),  the electronic spin feels the SG field combined with the effective magnetic field due to nuclei rotations. The latter field is along the $N$-vector, i.e. perpendicular to the molecular axis. A strong enough vertically polarized laser pulse excites molecular rotations in the vertical planes containing the $z$-axis. As a result, the rotation-induced effective magnetic field is perpendicular to the vertical SG field. Therefore, the force felt by the molecules is given by

\begin{equation}
Force=\frac{K_0B}{\sqrt{B^2+K_1^2}},\label{Fit to curve}
\end{equation}
where $K_0$ and $K_1$ are constants (the latter is proportional to $N$ or $P$).

Fig. \ref{LamdaSmall} presents results of the exact quantum-mechanical calculation of the SG force for our  imaginary $\widetilde{O_2}$ molecule. We considered the upper curves in  this figure, and tried to fit them to the above analytical expression.  We find an excellent agreement between the original data and the fitted curves, and the results of the fit are $K_1=0.61,0.25$ for $N=71,31$, respectively. We also find a  good agreement between the ratio of $N's$ (i.e., $71/31=2.3$) and of $K_1's$ (i.e., $0.61/0.25=2.4$).

\begin{figure}[htb]
\begin{center}
\includegraphics[width=70mm]{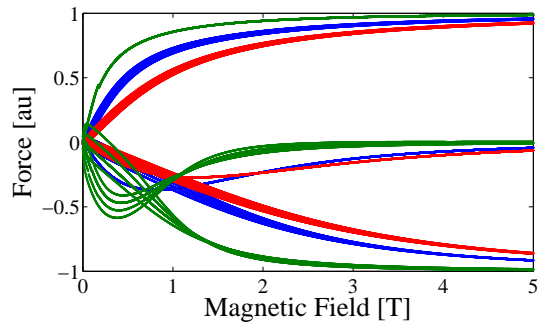}
\end{center}
\caption{Forces vs magnetic field for the $NH$ molecule (including the fine and hyperfine details), for $N=10$ (green), $N=30$ (blue) and $N=41$ (red). Only $M_F=1/2$ is considered here, but $M_F=-1/2$ gives the same results (only higher $M_F$'s will modify the spectrum). For $N=10$ the upper/middle/lower curves correspond to $J=N,J=N-1,J=N+1$, respectively, as in the Oxygen case. For $N=30,41$ the upper/middle/lower curves correspond to $J=N-1,J=N,J=N+1$, respectively, as in Fig. \ref{LamdaSmall}.}
\label{NH_forces}
\end{figure}

 As we have mentioned before, the $NH$ molecule is also characterized by a reduced value of  the spin-spin interaction compared to the spin-rotation interaction. Therefore, its dynamics should be  closer to the imaginary $\widetilde{O_2}$ molecule than to the real $O_2$ molecule considered above. In Fig. \ref{NH_forces} we plot some forces vs. the magnetic field for the $NH$ molecule, and indeed, observe a triplet-like structure similar to that one in Fig. \ref{LamdaSmall} (with the curve for $J=N$ being in the middle for large enough $N$).

By analyzing the results shown in Fig. \ref{NH_forces} we conclude that  the hyperfine structure details in $NH$ are not crucially important for our considerations, but the reduced spin-spin to spin-rotation interaction ratio for $NH$ defines the major difference of the deflection dynamics compared to the case of $O_2$. Also one notices the scaling with the magnetic field: here the higher values of the magnetic field are required ($2T$) to observe the  collapse of the broad distribution of forces into three narrow groups. This is due to the increased spin-rotation interaction for $NH$ (as compared to $O_2$). Fig. \ref{NH_forces} explains the behavior of the  distribution in Fig. \ref{NH_distribution}, where we have noticed that the three peaks are shifted as one increases the laser pulse strength.

\section{Conclusions} \label{Summary}
We considered scattering of paramagnetic molecules by inhomogeneous magnetic field in a Stern-Gerlach-type experiment. We showed that by prealigning the molecules before they interact with the magnetic field, one obtains an efficient control over the scattering process. Two qualitatively different effects were found, depending on the Hund's coupling case of the molecule. For molecules that belong to the Hund's coupling case (a), we showed that the deflection process may be strongly suppressed by laser pulses. This may be implemented as an optical switch in the molecular magnetic deceleration techniques \cite{even}. Furthermore, for the Hund's coupling case (b) molecules, a sparse distribution of the scattering angles is transformed into a distribution with several compact deflection peaks having controllable positions. Each peak corresponds to a scattered molecular sub-beam with increased brightness. The molecular deflection is considered as a promising route to the separation of molecular mixtures. Narrowing and displacing scattering peaks may substantially increase the efficiency of separating multi-component beams, especially when the prealignment is applied selectively to certain molecular species, such as specific isotopes \cite{isotopes}, or nuclear spin isomers \cite{isomers1,isomers2}.  One may envision more sophisticated schemes for controlling molecular scattering, which involve multiple pulses with variable polarization  for preshaping molecular angular distribution.   In particular, molecular rotation may be confined to a certain plane by using the "optical molecular centrifuge" approach \cite{centrifuge,Mullin}, double-pulse ignited "molecular propeller" \cite{propeller}, or permanent planar alignment induced by a pair of delayed perpendicularly
polarized short laser pulses \cite{France1,France2}. If molecules are prepared like this, a narrow angular peak is expected in their scattering distribution from a magnetic field. The position of the peak is controllable by inclination of the plane of rotation with respect to the deflecting field, similar to a related effect for molecular scattering in inhomogeneous electric fields see \cite{gershnabel4}). Moreover, further manipulations of the deflection process may be considered, e.g., by using several SG fields with varying directions. Magnetic deflection of $O_2$ molecules subject to laser-induced field-free manipulations, is currently a subject of an ongoing experimental effort.

\section*{ACKNOWLEDGMENT}

We enjoyed many stimulating discussions with Valery Milner and Sergey Zhdanovich. One of us (IA) appreciates the kind  hospitality and support at the University of British Columbia (Vancouver). This work is supported in part by grants from the Israel Science Foundation, and DFG (German Research Foundation).  Our research is made possible in part by the historic generosity of the Harold Perlman Family. IA is an incumbent of the Patricia Elman Bildner Professorial Chair.

\section{Appendix: $NH$ (Hund's coupling case (b)) Hyperfine structure} \label{NH HFS}
The hyperfine structure for the $NH$ molecule is described by the following Hamiltonian \cite{Carington,Klaus,Jesus,Lewen}:

\begin{eqnarray}
H_{HFS}&=&\sum_k {b_{F_k} T^1(\textbf{I}_k)\cdot T^1(\textbf{S})}\nonumber\\
&-&\sum_k{t_k\sqrt{10}T^1(\textbf{I}_k)\cdot T^1(\textbf{S},\textbf{C}^2(\omega))}\nonumber\\
&-& eT^2(\nabla\textbf{E}_2)\cdot T^2(\textbf{Q}_2)\nonumber\\
&+&\sum_k c_I(k) T^1(\textbf{N})\cdot T^1(\textbf{I}_k),
\label{hfs hamiltonian}
\end{eqnarray}
where the sum over $k=1,2$ represents the terms for both nuclei. The first term is the Fermi contact interaction, the second term is the dipolar interaction, the third one is the quadrupole term (this term exists only for the $^{14}N$), and the last term accounts for the nuclei spin-rotation interaction.
In the calculation of matrix elements we first coupled $\textbf{J}=\textbf{S}+\textbf{N}$, $\textbf{F}_1=\textbf{I}_H+\textbf{J}$ and only then $\textbf{F}=\textbf{I}_N+\textbf{F}_1$. All the matrix elements are diagonal in $F$, and the first three terms are given in \cite{Carington}.

The nuclear spin-rotation interactions are given by:

\begin{eqnarray}
&&\langle \eta,\Lambda,N,S,J,I_1,F_1,I_2,F,M_F|T^1(\textbf{N})\cdot T^1(\textbf{I}_1) \nonumber\\
&&|\eta,\Lambda,N',S,J',I_1,F_1',I_2,F,M_F\rangle\nonumber\\
&=& (-1)^{J'+F_1+I_1}\delta_{F_1,F_1'}\left\{ \begin{array}{ccc} I_1 & J' & F_1 \\ J & I_1 & 1 \end{array} \right\}\nonumber\\
&\times&\sqrt{I_1(I_1+1)(2I_1+1)}\delta_{N,N'}(-1)^{J'+N+1+S}\nonumber\\
&\times&\sqrt{(2J+1)(2J'+1)}\left\{ \begin{array}{ccc} N' & J' & I_1 \\ J & N & 1 \end{array} \right\}\nonumber\\
&\times&\sqrt{N(N+1)(2N+1)},\label{spin-rotation1}
\end{eqnarray}

where $I_1\equiv I_H$, and

\begin{eqnarray}
&&\langle \eta,\Lambda,N,S,J,I_1,F_1,I_2,F,M_F|T^1(\textbf{N})\cdot T^1(\textbf{I}_2) \nonumber\\
&&|\eta,\Lambda,N',S,J',I_1,F_1',I_2,F,M_F\rangle\nonumber\\
&=& (-1)^{F_1'+F+I_2}\left\{ \begin{array}{ccc} I_2 & F_1' & F \\ F_1 & I_2 & 1 \end{array} \right\}\nonumber\\
&\times&\sqrt{I_2(I_2+1)(2I_2+1)}(-1)^{F_1'+J+1+I_1}\nonumber\\
&\times&\sqrt{(2F_1+1)(2F_1'+1)}\left\{ \begin{array}{ccc} J' & F_1' & I_1 \\ F_1 & J & 1 \end{array} \right\}\nonumber\\
&\times&(-1)^{J'+N+1+S}\sqrt{(2J+1)(2J'+1)}\left\{ \begin{array}{ccc} N' & J' & S \\ J & N & 1 \end{array} \right\}\nonumber\\
&\times&\delta_{N,N'}\sqrt{N(N+1)(2N+1)},\label{spin-rotation2}
\end{eqnarray}
where $I_2\equiv I_N$. The constants were taken from \cite{Jesus}.
In the Zeeman Hamiltonian we consider only the contribution due to electronic spin:
\begin{equation}
H_{Z}=\mu_B g_s T^1(\textbf{B})\cdot T^1(\textbf{S}), \label{ZeemanHfs}
\end{equation}
and we neglect other small contributions coming from the nuclei's rotation and spin, and electronic anisotropy. The Zeeman matrix element is proportional to:
\begin{eqnarray}
&&\langle \eta,\Delta,N,S,J,I_1,F_1,I_2,F,M_F|T^1_0(\textbf{S})\nonumber\\
&&|\eta,\Delta,N,S,J',I_1,F_1',I_2,F',M_F\rangle\nonumber\\
&=& (-1)^{F-M_F} \left( \begin{array}{ccc} F & 1 & F' \\ -M_F & 0 & M_F \end{array} \right)\nonumber\\
&\times& (-1)^{F'+F_1+1+I_2}\sqrt{(2F+1)(2F'+1)}\left\{ \begin{array}{ccc} F_1' & F' & I_2 \\ F & F_1 & 1 \end{array} \right\}\nonumber\\
&\times& (-1)^{F_1'+J+1+I_1}\sqrt{(2F_1+1)(2F_1'+1)}\left\{ \begin{array}{ccc} J' & F_1' & I_1 \\ F_1 & J & 1 \end{array} \right\}\nonumber\\
&\times&(-1)^{J'+S+1+N}\sqrt{(2J'+1)(2J+1)}\left\{ \begin{array}{ccc} S & J' & N \\ J & S & 1 \end{array} \right\}\nonumber\\
&\times&\sqrt{S(S+1)(2S+1)},\label{hfs Zeeman}
\end{eqnarray}
where it is no more diagonal in F.
Finally, for the alignment calculations the following matrix element is useful:

\begin{eqnarray}
&&\langle \eta,\Delta,N,S,J,I_1,F_1,I_2,F,M_F|D^{2 *}_{00} \nonumber\\
&&|\eta,\Delta,N',S,J',I_1,F_1',I_2,F',M_F\rangle\nonumber\\
&=& (-1)^{F-M_F} \left( \begin{array}{ccc} F & 2 & F' \\ -M_F & 0 & M_F \end{array} \right)\nonumber\\
&\times& (-1)^{F'+F_1+2+I_2}\sqrt{(2F+1)(2F'+1)}\left\{ \begin{array}{ccc} F_1' & F' & I_2 \\ F & F_1 & 2 \end{array} \right\}\nonumber\\
&\times& (-1)^{F_1'+J+2+I_1}\sqrt{(2F_1+1)(2F_1'+1)}\left\{ \begin{array}{ccc} J' & F_1' & I_1 \\ F_1 & J & 2 \end{array} \right\}\nonumber\\
&\times&(-1)^{J'+N+2+S}\sqrt{(2J+1)(2J'+1)}\left\{ \begin{array}{ccc} N' & J' & S \\ J & N & 2 \end{array} \right\}\nonumber\\
&\times&(-1)^N \left( \begin{array}{ccc} N & 2 & N' \\ 0 & 0 & 0 \end{array} \right) \sqrt{(2N+1)(2N'+1)}.\label{Prealignment HFS}
\end{eqnarray}

\bibliographystyle{phaip}

\end{document}